# Dynamic optical properties of metal hydrides

Kevin J. Palm[1,2], Joseph B. Murray[1], Tarun C. Narayan[1], and Jeremy N. Munday[1,3,a)]

[1]Institute for Research in Electronics and Applied Physics, University of Maryland, College Park, Maryland 20742, USA
[2]Department of Physics, University of Maryland, College Park, Maryland 20742, USA
[3]Department of Electrical and Computer Engineering, University of Maryland, College Park, Maryland 20742, USA
[a)]Authors to whom correspondence should be addressed: jnmunday@umd.edu

## Abstract

Metal hydrides often display dramatic changes in optical properties upon hydrogenation. These shifts make them prime candidates for many tunable optical devices, such as optical hydrogen sensors and switchable mirrors. While some of these metals, such as palladium, have been well studied, many other promising materials have only been characterized over a limited optical range and lack direct *in situ* measurements of hydrogen loading, limiting their potential applications. Further, there have been no systematic studies that allow for a clear comparison between these metals. In this work, we present such a systematic study of the dynamically tunable optical properties of Pd, Mg, Zr, Ti, and V throughout hydrogenation with a wavelength range of 250 - 1690 nm. These measurements were performed in an environmental chamber, which combines mass measurements via a quartz crystal microbalance with ellipsometric measurements in up to 7 bar of hydrogen gas, allowing us to determine the optical properties *during* hydrogen loading. In addition, we demonstrate a further tunability of the optical properties of titanium and its hydride by altering annealing conditions, and we investigate the optical and gravimetric hysteresis that occurs during hydrogenation cycling of palladium. Finally, we demonstrate several nanoscale optical and plasmonic structures based on these dynamic properties. We show structures that, upon hydrogenation, demonstrate five orders of magnitude change in reflectivity, resonance shifts of >200 nm, and relative transmission switching of >3000%, suggesting a wide range of applications.

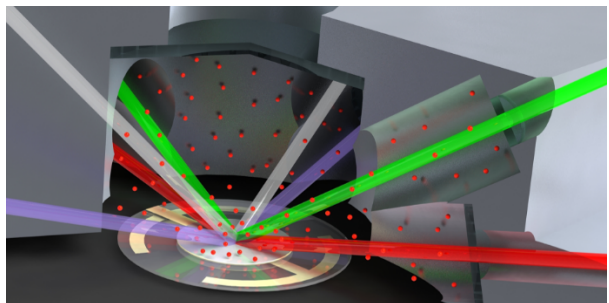
TOC Graphic

# Introduction

Materials with tunable optical properties are critical to the development of novel active nanophotonic devices ranging from plasmonic light absorbers and biosensors to switchable mirrors.[1–8] The ability to change the resonances of a structure *in situ* allows for increased functionality and enables new device architectures. One particular class of materials well-suited for tunable applications is metal hydrides. A number of metals have been shown to strongly absorb hydrogen, resulting in hydrogen to metal atom ratios approaching or even exceeding 1:1.[9–15] These metals typically undergo crystalline phase transitions, altering their crystal and electronic structures. The large changes to the crystal structure, the additional electrons, and the additional resonances associated with the binding energy of the hydrogen to the lattice can create dramatic shifts in the metal's optical properties.

These metal hydrides are of great interest for switchable photonic devices, particularly for applications involving optical hydrogen sensors and switchable mirrors. Palladium and palladium alloys have been widely used for such sensors, structured as both thin films and nanoparticles.[16–23] Yttrium and lanthanum have been investigated for their use as switchable mirrors due to their metal to dielectric transition upon hydrogenation.[7,8] Magnesium has seen recent interest for use in reversible color filters due to its optically dramatic shift from metal to dielectric upon hydrogenation.[24] Hafnium has been introduced as an optical hydrogen sensor that can span six orders of magnitude.[25] Niobium nanorods were recently investigated as a new material for high temperature plasmonics with switchable properties upon hydrogenation.[26]

While work has been done on the optical properties of metals and their hydrides, these previous studies have a variety of limitations, including: narrower wavelength ranges (250 - 1690 nm in this work), lack of any dynamic or intermediate hydride data, or temperature ranges that prevent comparison across different studies. On top of these limitations, varied fabrication conditions and procedures alter the optical properties of a metal in a given experiment, furthering the difficulty of comparison. No systematic work has compared a wide range of materials (here Pd, Mg, Zr, Ti, and V) prepared and tested under identical broadband illumination and time-dependent hydrogenation conditions. Further, none of referenced studies include direct *in situ* loading measurements (Azofeifa *et al.* indirectly infer loading *in situ* via resistivity and transmission spectra)[27].

By addressing all of the above issues in this work, we present a complete, uniform set of measurements that can be used to compare the optical properties of five different metal thin films (Pd, Mg, Zr, Ti, and V) before, during, and after hydrogenation. We pair this data with simultaneously recorded measurements of the metal/hydrogen atom ratio using a custom environmental chamber incorporating a quartz crystal microbalance (QCM).[28] We also investigate the effects of annealing on the optical properties of Ti and TiH$_x$ to quantify its strong dependence on preparation conditions, which offers another knob for optical response tunability. We perform a cycling experiment on Pd to study the hysteresis between hydrogenation cycles and to determine the correlation of the loading value with cycled optical properties. Finally, we demonstrate the applicability of the tunable optical properties of these metal hydrides in dynamically controlled nanostructures and thin film cavities. We find that the dramatic changes in the optical response of these materials with the hydrogenation reaction presents a wealth of possibilities for practical devices.

## Results and discussion

Characterizing the optical properties of hydrides while simultaneously and independently measuring the loading introduces several challenges to the standard ellipsometric scheme. To overcome these challenges, we perform measurements in a modified version of our environmental chamber described in Murray *et al.*, which maintains the sample at a constant pressure and temperature while varying the hydrogen partial pressure.[28] This system monitors the mass change of thin films via a QCM. A specially designed ellipsometry lid provides optical access to the sample with fused silica windows set normal to our desired optical measurement angles to prevent extraneous Fresnel effects (polarization dependent transmission magnitude and phase) in the measurement (see Methods for further details on the ellipsometry measurement scheme, accounting for stress-induced residual retardation in the optical measurement, and correcting for artifacts in QCM measurements).

Figure 1 presents the dielectric functions of five metals (Pd, Mg, Zr, Ti, and V) as they hydrogenate over the entire visible to near-infrared spectral range, 250 to 1690 nm. The bottom panels show the hydrogen loading (*i.e.* number of hydrogen atoms per metal atom in the lattice) as a function of time, and the colored dots represent points in time where the optical properties are recorded (bottom plots in each panel). Measurements were taken on ~200 nm thick metal films capped with 3 nm of Pd that were annealed for 2 hours at 350 °C under < 1 mtorr vacuum (with the exception of Mg, which had a thickness of 25 nm). During Mg hydrogenation, the Mg closest to the Pd capping layer hydrogenates, turning primarily to $MgH_2$. This $MgH_2$ layer is a poor proton conductor and acts as a blocking layer for more hydrogen to penetrate into the film, causing the time scale for total, bulk hydrogenation to be several days.[29] Because of this phenomena, we performed measurements on a 25 nm Mg thin film (the thickness of the formed blocking layer) in order to fully hydride the sample. The Pd capping layer is necessary for each metal (other than Pd itself) in order to reduce the activation energy of $H_2$ splitting and allow for diffusion into the bulk.[30] This capping layer also prevents oxidation of the underling film, keeping it pristine. Note that in the case of the $V/VH_x$ data, wavelengths below 300 nm are not available due to high stress-induced ellipsometric retardation (*i.e.* polarization dependent phase change of the transmitted light) in the windows during this measurement (see Methods).

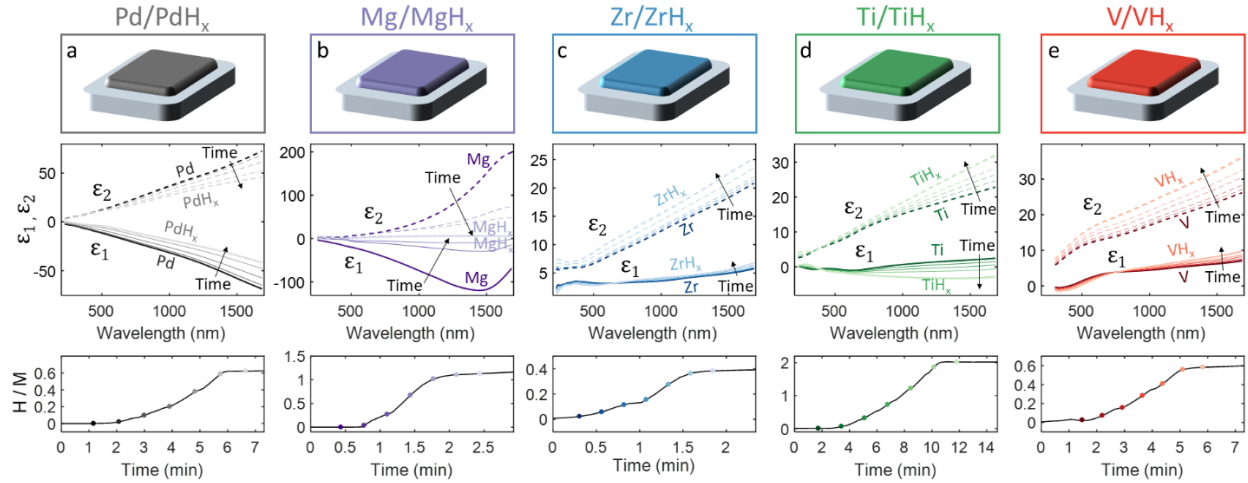

**Figure 1**: Dynamic optical properties and loading measurements of Pd, Mg, Zr, Ti, and V metals and their hydrides. The top panel shows schematics of the thin metal films used for the measurements (note: the 3 nm Pd capping layer used in the experiment is not shown). The middle panel displays the dielectric function of each metal and its hydride, as well as the intermediate loading states. The dynamic loading data is shown in the bottom panel where H/M is the number of hydrogen atoms per metal atom in the metal lattice. The colored dots indicate the times corresponding to the optical measurements in the graphs. $\varepsilon_1$ is shown as solid lines and $\varepsilon_2$ as dashed lines. As time progresses, the shading of the lines gets lighter.

Figure 2 shows the dependence of the optical properties on the measured loading values of the metals. Two main points can be drawn from this data. First, Mg's dramatic change in optical properties is clearly evident in its dependence on the loading values. Second, Ti has the most interesting relationship because the real part of the dielectric function has a different slope depending on the wavelength of illumination. The feature could enable a resonance to shift to longer or shorter wavelengths upon hydrogenation depending on the incident wavelength.

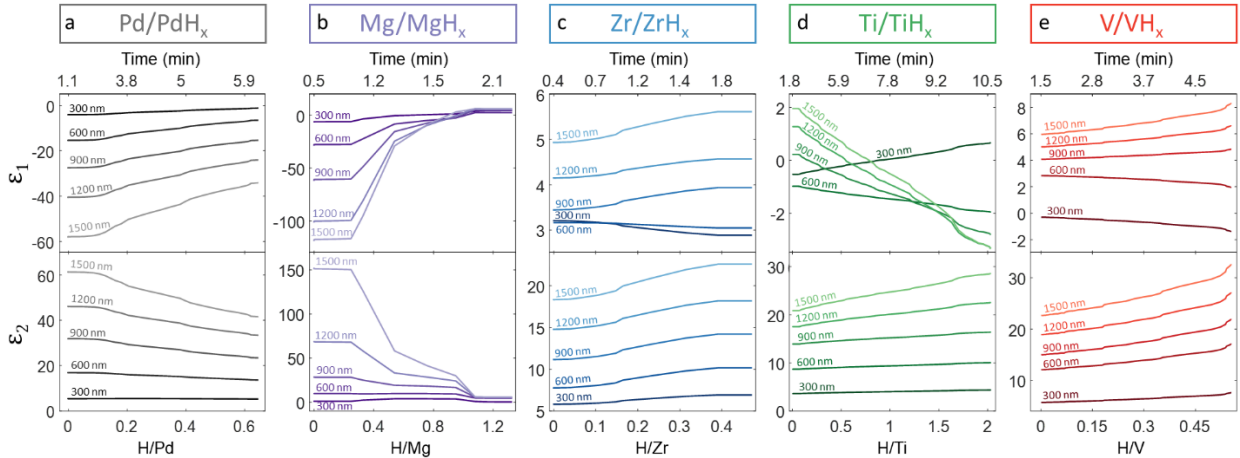

**Figure 2:** Optical properties of the metal hydrides as a function of hydrogen to metal ratio. Ti is of particular interest because the slope of the real part of the dielectric function with hydrogen loading depends on the wavelength investigated. This may present a new scheme for optical detection of hydrogenation in Ti. Note that the upper time axis is not linearly spaced.

The results of all of the optical and loading measurements for the five different metals and their hydrides are discussed in turn below.

*Pd/PdH$_x$*
The optical properties of both the pure Pd metal and the PdH$_x$ films agree well with literature values across the measured range, which serves as a standard for this measurement technique.[31,32] There are no clear peaks or inflection points in the Pd film's optical properties within the wavelength range under investigation. Upon hydrogenation, Pd becomes distinctly less metallic at a fairly uniform rate with the real part of the dielectric function increasing by up to 39% in the near infrared, and the imaginary part decreasing by similar percentages. The film reaches a final loading value of 0.67 ± 0.12, where the loading value is defined as the number of hydrogen atoms per metal atom in the lattice, in agreement with previous results.[10,11]

*Mg/MgH$_x$*
Of the metals investigated, Mg has the most dramatic optical changes. The optical properties of the metal, which matches well with reported data over the previously studied wavelength region[33,34], is the most lossy, with the imaginary part of the dielectric function three times greater than that of the next metal, Pd. During hydrogenation, it can be seen that while mass loading proceeds continuously, the optical properties change abruptly from a metal to an insulator, which is consistent with previous electrical measurements.[35] When comparing the dielectric function of MgH$_x$ with those reported in literature, we find that our measured film appears more metallic (characterized by a decrease in the real part and increase in the imaginary part of the dielectric function with wavelength). One likely cause for this discrepancy is that Mg and Pd readily form an alloying layer.[36,37] This alloy does not become entirely a dielectric, as we expect for MgH$_2$, which would cause our film to appear more metallic than the pure MgH$_2$. After accounting for apparent oxide formation (see Methods), the Mg demonstrated loading of 1.3 ± 0.5, which is lower than the expected 2.0 loading value. The alloying layer discussed above could account for this lower measured value. Another potential cause could be that the bottom few nm of the film did not fully hydride due to the blocking layer.

*Zr/ZrH$_x$*

Zirconium has the smallest change in optical properties of the metals being investigated. The real part of the dielectric function exhibits very small change during the loading process, but the imaginary part increases significantly. The Zr film also exhibited the lowest amount of loading amongst the films, only reaching a value of 0.47 ± 0.05. The Zr and ZrH$_x$ optical data shown here differ from the values reported by Azofeifa *et al.*[38] (note that those values also greatly deviate from previous data for Zr[39]). Thus, there appears to be an important factor in sample preparation yet to be fully described. However, the magnitude of our measured shift in the dielectric function upon hydrogenation agrees very well with the data presented by Azofeifa *et al.* over the wavelength range they explored.[38] The difference in initial properties could be attributed to metal preparation conditions, as the grain size of the metal has an effect on the optical properties of the hydrides.[32] We also observe that at 640 nm, our data shows that there is zero change in the real dielectric function upon hydrogenation. Points like these may be useful as reference points in a differential measurement scheme.

*Ti/TiH$_x$*

The derived optical properties of the initial, pristine Ti metal are in good agreement with previously reported data over the same wavelength range measured here[40]. The local maxima in the real part of the dielectric function at ~400 nm and ~800 nm, typically ascribed to d-band transitions[41,42], are also clearly visible. After hydrogenation, these undulations disappear and the TiH$_x$ appears to become more metallic, characterized by a decrease in the real part and an increase in the imaginary parts of the dielectric function with wavelength. However, the previous studies have not observed an increase in Ti conductivity upon hydrogenation, which would have explained this behavior.[43] This phenomenon may simply be due to the elimination of the above mentioned d-band oscillations. Also, a small, broad peak at ~250 nm is visible in the data, which is characteristic of the metal-hydrogen bond.[44] Ti loads to the highest value of any metal measured in this experiment, achieving a value of 2.04 ± 0.12.

*V/VH$_x$*

Over the visible wavelength range previously reported, our V metal data exhibits similar trends.[27] Band structure theory predicts that there should be two absorption peaks in the data centered at 406 and 708 nm.[45] In the literature, it has been observed that thermal broadening merges these peaks into one broad peak. Azofeifa *et al* found this peak to be at 520 nm, and we find this peak at a similar location of 510 nm. The broadening of this peak is attributed to large electron lifetimes in the excited states.[46] Upon hydrogenation, we find that V has a small change in the real part of the dielectric function with a much more significant change to the imaginary part, which increases by more than 38% in the near infrared region. Further, we find that at 750 nm, there is no change in the real part of the dielectric function even though the imaginary part changes significantly upon hydrogenation, showing that the imaginary part of the dielectric function can be controlled independently from the real part through hydrogenation over this bandwidth. This phenomenon was also observed previously, but near 500 nm. We attribute this difference to sample preparation and the resulting difference in the grain size of the metal. The V film achieves a loading of 0.56 ± 0.03.

*Effect of Ti Annealing*
We found that the Ti and TiH$_x$ optical properties were quite sensitive to preparation conditions. To further explore this phenomenon, we varied the annealing conditions. Ti samples were either not annealed, annealed at 200 °C for 2 hours, or annealed at 350 °C for 2 hours, with each anneal occurring in under <1 mtorr vacuum. These samples were then characterized using the same process described above. Figure 3 shows the measured dielectric functions. This experiment revealed two interesting characteristics of Ti. When not annealed, the oscillations associated with d-band transitions mentioned above are clearly visible and become less prominent with increased annealing temperature. The hydrogen loading generally reduces the impact of annealing, as the hydrides exhibit a smaller optical property change from annealing than the pure metals. There are a few possible explanations for this effect. First, the very large stresses involved in creating the hydride may be producing enough dislocations that it is partially undoing the impact of annealing. Second, it is known that the effect of defects on resistivity is reduced with higher hydrogen concentration and perhaps this is the effect we observe; however, additional studies are needed to understand the root cause.[43] Nevertheless, it is clear that annealing offers an additional opportunity for modifying how hydrogenation affects the optical properties of Ti.

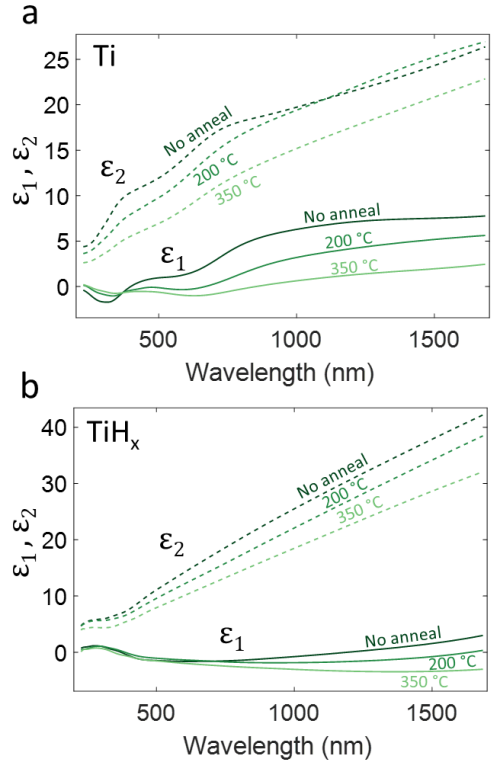

**Figure 3**: Effects of annealing on the optical properties of Ti and TiH$_x$. (a) Dielectric function of Ti with different levels of annealing. (b) Dielectric function of TiH$_x$ measured on the same samples.

## Pd hysteresis

Some devices, such as hydrogen sensors, require the ability to cycle the metal between a hydrogenated and unhydrogentated state. Palladium is unique compared to the other metals because it unloads at room temperature with easily realizable, low hydrogen partial pressures, allowing cycling by simply varying the hydrogen gas content in the chamber. We use this process to measure the optical properties and the loading of the Pd through three consecutive loading cycles, shown in Figure 4. The red shaded regions in Figure 4a indicate when H$_2$ is flowing into the chamber at 20 sccm and hydrogen is loading the metal film. The blue shaded regions indicate unloading phases where Ar fills the chamber, displacing the H$_2$, and the metal film desorbs.

We can see in Figure 4a that there is a clear hysteresis in the optical properties of the Pd metal between the first loading cycle and the subsequent measurements. However, after the Pd had been loaded once, there is no discernible difference in optical properties between the Pd after the first and the second unload, eliminating further hysteresis. Note that there is no discernible hysteresis in the optical response of the hydride, as demonstrated in Figure 4c. More data would be needed to determine performance over a longer timescale (or number of cycles), but the data imply the optical hysteresis of Pd can be greatly reduced with an initial treatment of the metal with H$_2$. Note that the above refers only to the inter-cycle hysteresis. The well-known intra-cycle hysteresis where the α to β phase transition happens at a different hydrogen partial pressure than the β to α phase transition would still be present.[9] This inter-cycle phenomenon is most likely attributed to deformations formed in the Pd lattice upon the first hydrogen loading. We observe that the loading value for each PdH$_x$ state remains consistent throughout all three cycles, although we observed and increase in unloading time between the first and the second cycles (see Methods for details on loading calculations).

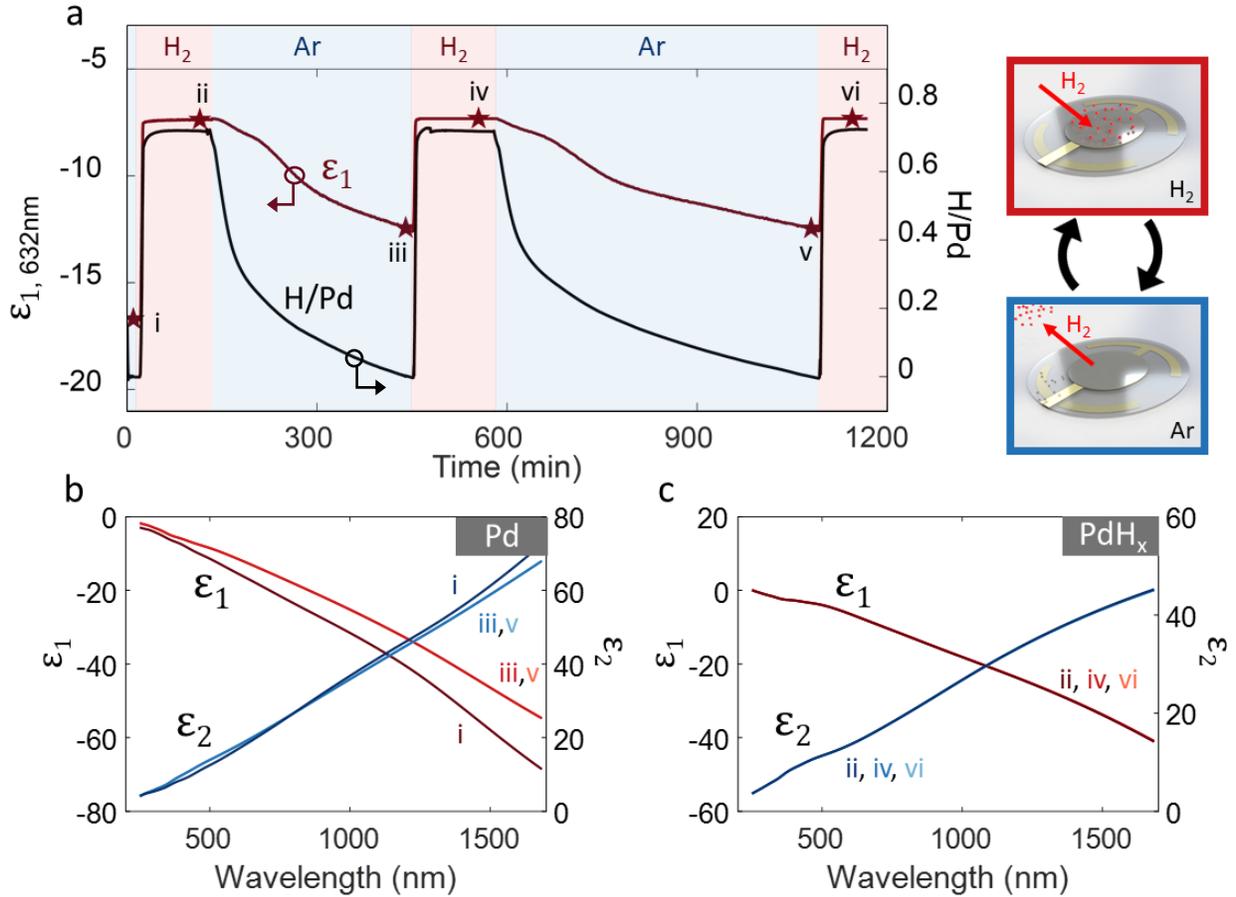

**Figure 4**: The optical and loading response of Pd during hydrogen cycling. (a) Dynamic loading data (black) and the real dielectric permittivity (red) at 632 nm plotted over three loading cycles of a 200 nm Pd film. The red shaded areas indicate periods of loading during which H$_2$ is flowing into the chamber, while the blue shaded sections depict times during which Ar is flowing to flush out the H$_2$. The stars on the plot indicate the times at which the ellipsometric measurements shown in (b) and (c) were made. (b) shows the Pd metal data and (c) shows the PdH$_x$ data. Note for the second two Pd measurements (iii and v) and all three PdH$_x$ measurements (ii, iv, vi) that the measured dielectric functions are indistinguishable and, thus, overlap each other in these plots.

# Tunable nanophotonics with metal hydrides

There are many potential applications presented by the tunable optical properties of the metal hydrides investigated in this work. In this section, we detail four examples of the potential uses of these materials: nano-particles with variable scattering cross sections, nanorod arrays with tunable transmission, thin film resonators with adjustable color filtering, and hydrogen switchable perfect absorbers.

We begin with the simplest of these examples, showing the ability to tune the cross section of nanoparticles in free space by hydrogenating the sample. The top panels in Figure 5 show a schematic of the configuration with the plane wave source being scattered by the metal and its hydride, depicting whether they primarily increase or decrease in total scattering upon hydrogenation. The bottom panels show the difference in the simulated finite difference time domain (FDTD) scattering cross sections between the metal and its hydride (see Figure S2 for total scattering cross sections). First, a clear distinction can be made between two types of the metals: Pd and Mg particles primarily decrease total scattering upon hydrogenation, while Zr, Ti, and V particles show mostly increased scattering. This effect is caused by opposite responses of the metal's index of refraction, with Pd and Mg decreasing and Zr, Ti, and V increasing. Mg clearly has the largest differential response to hydrogenation, having an order of magnitude larger change than the other metals. This was expected due to its dramatic metal to insulator transition. The Pd nanoparticles have the next largest differential response and behave similarly to Mg, having a large change in response in the visible to ultraviolet range.

Ti demonstrates wavelength-dependent regions of both increased and decreased scattering cross section – increasing near ultraviolet scattering while decreasing near infrared scattering. It also is the only metal under investigation that has its largest scattering change in the near infrared range (400 nm particle diameter), presenting applications that the other metals could not provide. However, Ti did exhibit a fairly small differential scattering response compared to the other metals. V primarily has its largest scattering changes in the ultraviolet. The exact magnitude and location of many of these peaks cannot be determined because the simulation was limited to the experimentally measured wavelength range (> 250 nm), which cut off these peaks. As expected, Zr has the smallest differential response because of its small change in optical properties upon hydrogenation. The response is on the same order as Ti, with its maximum responses spanning a wider range of wavelengths than the other materials, beginning in the ultraviolet and continuing to the near infrared.

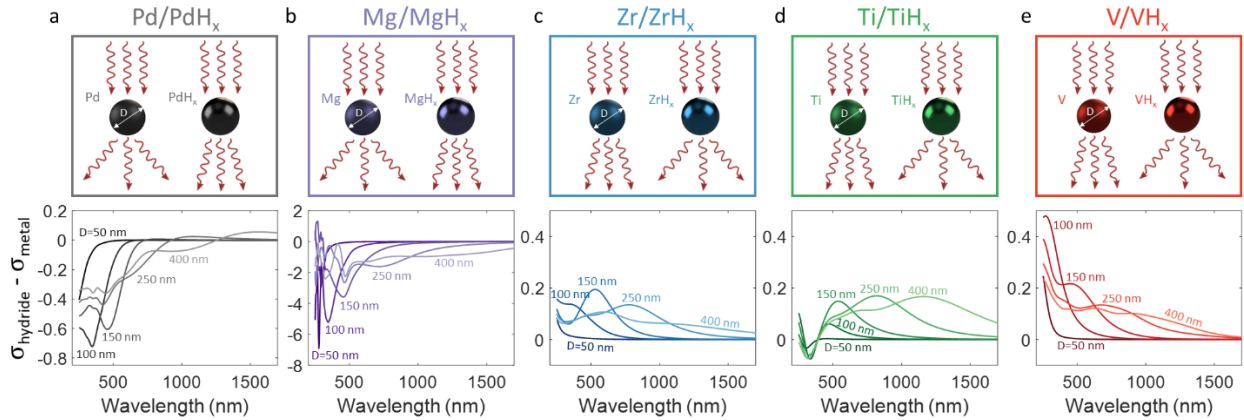

**Figure 5**: Differential scattering cross sections for nanoparticles in free space composed of different metals. The top panels show schematics illustrating the heuristic change in the nanoparticle cross section resulting from the change in the dielectric response. The bottom panels show the differences in the scattering cross sections of metals and their hydrides for multiple particle diameters ranging from 50 - 400 nm. For Pd and Mg, hydrogenation causes decreased scattering, as opposed to Zr, Ti, and V where the scattering increases.

Next, we simulate periodic arrays of nanorods on a glass substrate, showing large resonance shifts and changes in transmission upon hydrogenation. The relative change in transmission upon hydrogenation for these metals is shown in Figure 6. The arrays have a 500 nm period in both the parallel and perpendicular directions, a rod width of 100 nm, and a length that varies from 150 to 400 nm (the electric field is parallel to the length of the rods). Once again, Mg and Pd behave differently than the other metals in terms of responses upon hydrogenation: Mg and Pd have an increase in transmission upon hydrogenation, while Zr, Ti, and V exhibit a decrease in transmission.

Mg has the strongest relative response, exhibiting a full 3800% relative transmission increase for the 400 nm rods. Pd shows the next highest response, with a 190% relative increase in transmission at the 400 nm length. Mg and Pd have the narrowest peaks, causing for a more spectrally localized response upon hydrogenation. Zr and V have very similar resonant response locations in the visible and near infrared, with V exhibiting stronger changes in transmission. These materials have their peak changes in the visible and near infrared. Ti has transmission magnitude shifts similar to Zr, but with its responses spanning further in the infrared, allowing for a wider usable bandwidth. These transmission differences allow for *in situ* tunability of optical responses for various wavelength ranges with without having to physically move any piece of the structure or electrically activate any part of it.

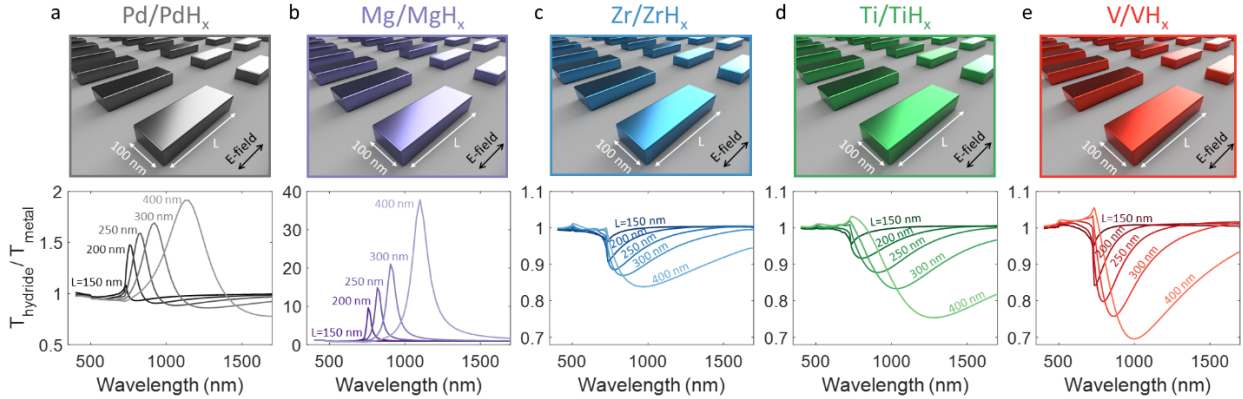

**Figure 6**: Relative change in transmission upon hydrogenation of periodic nanorod arrays. The top panels show schematics of a nanorod array on a glass substrate. The rods are spaced 500 nm apart in the both the parallel and perpendicular directions and are 100 nm wide. The rod length is varied from 150 to 400 nm. The polarization of the electric field is parallel to the length of the rods. The bottom panels show the relative difference of the transmission spectra between the metals and their hydrides. The transmission for Mg and Pd nanorods increases upon hydrogenation for most of the spectrum, while those made of Zr, V, and Ti decrease.

The dynamic optical properties demonstrated above suggest themselves to two potential thin film applications: tunable color filters and switchable perfect absorbers. Figure 7 depicts two embodiments of these devices: a Fabry-Perot cavity with the metals acting as both the top (partial) mirror and the bottom mirror separated by a dielectric and a thin film Si absorber with the metals acting as a lossy bottom mirror. For each metal, several representative thickness of the Si or $SiO_2$ layers are shown. Note that tunable color filters and switchable perfect absorbers are not mutually exclusive and differ only in their figure of merit, allowing a single structure to act as both. In all respects the Fabry-Perot cavity has better performance. However, the Si systems are of interest due to their simplicity and because a Si based perfect absorber is useful for several applications, such as in a Si/metal photodetector.[47] We present here only the materials that function well as both tunable filters and switchable perfect absorbers in either structure: Mg, Pd, and Ti.

Tunable color filtering is characterized by shifts in the wavelength of the resonance.[24] In each case shown, hydrogenation of the metal produces an easily measurable and generally visible change in the resonant wavelength. The largest shifts in the Fabry Perot modes shown are 360 nm (or 1 eV), 100 nm (or 0.4 eV), and 50 nm (or 0.1 eV), for the Mg, Pd, and Ti, respectively. The largest shift in the resonances of Si on metal are 40 nm (or 0.24 eV), 50 nm (or 0.17 eV), and 70 nm (0.04 eV) for the Mg, Pd, and Ti, respectively. These values compare well with Duan *et al.* which demonstrated wavelengths shifts of ~150 nm in a Mg nanoparticle system or wavelength shifts of ~150 nm for Pd nanorings presented by Zorić *et al*.[24,48]

On the other hand, the figure of merit for the switchable perfect absorbers is the difference in maximum absorption or reflection.[49–51] The six structures shown here also demonstrate excellent performance as switchable perfect absorbers. In fact, all of the systems in Figure 7 have resonances that have more than 2 order of magnitude of switchability, including extreme cases like the Pd Fabry-Perot structure with 121 nm of SiO$_2$, which changes absorption by more than five orders of magnitude, or the Mg Fabry Perot structure with 204 nm of SiO$_2$, which switches by more than four orders of magnitude. This offers significantly improved performance compared to Walter *et al.* who presented reflection changes in Pd nanodiscs or Tittl *et al.* who demonstrated perfect absorbers in Pd gratings, who both reported reflection ratios of <1000.[50,51] As above, it is interesting to note that the structures involving Ti exhibit unique behavior amongst the materials we investigated. The resonances in the Ti devices shift to shorter wavelengths after the hydrogenation reaction due to the increase in refractive index, in contrast to the red shifts demonstrated with the other two metals. Using this effect, it may be possible to combine layers of, for example, Pd and Ti to exaggerate resonant shifts. These types of simple resonant structures show the great promise for materials as a part of tunable optical systems.

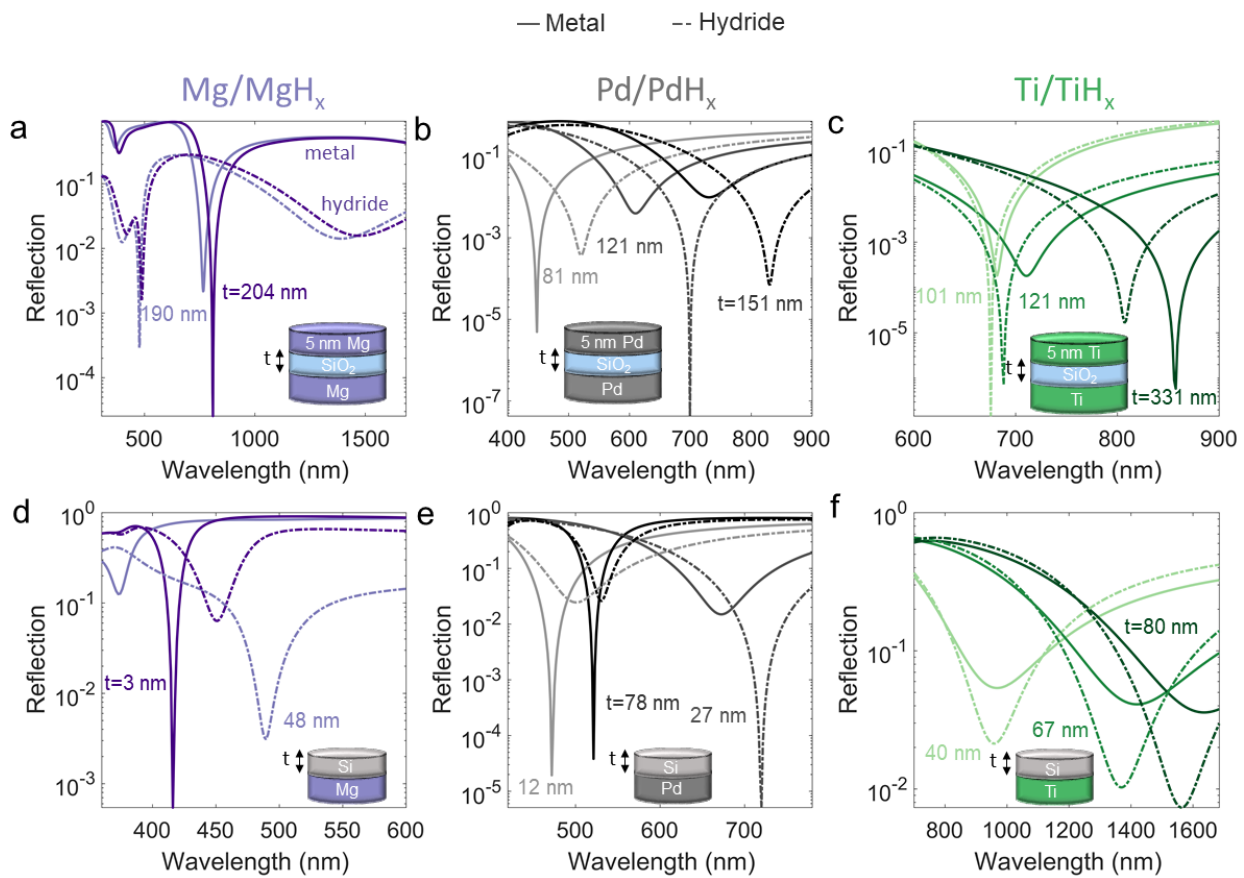

**Figure 7**: Switchable perfect absorbers and tunable color filters using thin film structures containing the three metals that offer the largest optical response to hydrogenation: Mg, Pd, and Ti. The thickness of either the Si or SiO$_2$ layers are labelled on the charts for each colored curve, and the bottom metal is considered bulk. The pure metal is denoted with a single solid line and the metal hydride with dashed lines. (a-c) Fabry-Perot cavities comprised of Mg, Pd, and Ti, respectively, for top and bottom mirrors separated by SiO$_2$. (d-f) Si thin film structures on Mg, Pd, and Ti, respectively, undergoing hydrogenation.

All of these structures perform well as either switchable perfect absorbers (>100x changes in reflectivity) or tunable color filters (a minimum of 40 nm resonant wavelength shift).

# Conclusions

In summary, we have mapped the dynamic optical properties of the hydrogenation reaction of five different metal films and have simultaneously recorded the real-time hydrogen loading data. We have shown that the optical properties can be tuned for the different metals, with the largest changes exhibited in Mg and the smallest changes in Zr. We demonstrated further potential tunability for Ti structures by determining the effect of annealing on optical property changes in the metal and its hydride. Pd was shown to have a large hysteretic optical response on the first cycle, and then very consistent optical properties between the second and third cycles, while having no hysteresis in the hydrided state. Four different nanostructured geometries were studied based on the measured optical properties, which demonstrated several potential applications for the tunability of these metals. Mg was shown to have the largest response for all structures due to its extreme change in optical properties upon hydrogenation. Mg and Pd structures exhibited a red-shift in their resonances, while Zr, V, and Ti systems are characterized by blue shifts. Ti, Mg, and Pd were all found to be excellent candidates for perfect absorbing devices, making them widely applicable for such designs. This work acts both as a point of comparison between these materials and to demonstrate their usefulness as tunable optical materials for novel dynamically switchable photonic devices.

# Methods

*Sample Fabrication*

The substrate for each of the samples is a 5 MHz Inficon QCM. Before sample deposition, each QCM is cleaned with acetone, methanol, isopropyl alcohol, and water to remove any particles or organics on the surface of the Au electrode. The 12.7 mm diameter thin film disks are defined by a machined Al shadow mask. Each metal is deposited as part of a stack using electron beam evaporation (Angstrom NEXDEP). First, a 3 nm Cr adhesion layer is deposited, followed by 200 nm of the metal under test (25 nm for Mg), and finished with a 3 nm Pd capping layer. This final layer acts as a semi-transparent and permeable surface to split the $H_2$ molecules. For each deposition, at least two QCMs are included: the first for the optical and *in situ* loading measurement and the second for a more precise and complete loading measurement in a separate environmental chamber which incorporates stress compensation (see Murray *et al*. for details).[28] Having the samples deposited in the same run ensures the similarity of the metals on each QCM for comparison of loading. With each deposition, a lithographically defined 1 cm x 1 cm square is included for determining the sample height via atomic force microscopy (AFM) (Cypher, Asylum Research). Immediately after sample deposition, the samples are annealed in < 1 mtorr vacuum at 350 °C for 2 hours with the exception of the Ti samples used in the annealing study.

*Optical Measurement*

The optical properties for each of the materials are measured via spectroscopic ellipsometry (Woollam M-2000). Each sample is measured immediately after annealing in order to avoid contamination of the sample or excessive oxidation. Before an optical measurement, the sample chamber is purged with Ar at ~200 sccms for a minimum of 1 hour to remove any trace hydrogen left in the system. Assuming complete mixing, this brings the $H_2$ partial pressure in the 45 cm$^3$ chamber to <10$^{-5}$ bar. Prior to being placed in the chamber, an optical measurement of the sample is taken (spectral range 193-1690 nm, angles varied from 50 to 75°). The sample is then mounted on the QCM stage in the environmental chamber and the chamber is pressurized to 7 bar in Ar. The optical properties are then recorded at the 4 separate inlet angles of the the environmental chamber (48°, 55°, 70°, 75°). These results are compared to the initial out-of-chamber measurement to determine the change in the phase difference between TE and TM polarizations as light passes through the chamber windows (ellipsometric retardation effects). The origin of this retardation is the birefringence in the glass produced by anisotropic window stress when mounting the chamber lid. A dynamic ellipsometric measurement is then taken through the 75° window (approximately 6 data points per minute) and the chamber flow rate is set to 20 sccm $H_2$. The dynamic measurement is stopped when both the optical properties and the measured frequency have stabilized. While still under $H_2$ flow, ellipsometric measurements are taken of the metal hydride using all 4 inlet angles. During the Pd cycling measurements the $H_2$ is flushed from the chamber by flowing 60 sccm of Ar during the unloading steps.

*Optical Fitting*

All ellipsometric data is fit using the Woollam CompleteEASE fitting software. The pure metal fitted optical parameters are created using a Kramers Kronig consistent B-spline to minimize the difference between the modelled ellipsometric data and the measured data. For each metal, the surface roughness was recorded before and after a loading run using AFM. The RMS surface roughness from these experiments is input into the optical model for that metal. The sample thickness is determined by step height measurements with the AFM on the lithographically defined Si square included with the sample deposition and is input into the model. The Pd capping layer thickness in the fit is constrained between 2-4 nm, with the thickness value being a fit parameter. The optical data used for the Pd capping layer on each sample is the measured optical properties of the Pd sample film shown in this manuscript.

To determine the ellipsometric retardation phase effects (given by the added TE/TM phase difference) of the environmental chamber windows, the pure metal model found above is fixed and the difference between the measured phase through each window and the phase of the pure metal model is fit using the following equation:

$$\Delta(f) = f(C_1 + C_2 f^2)$$

where $\Delta$ is the frequency dependent retardation effects input to the model, $f$ is the optical frequency of the spectroscopic beam, and $C_1$ and $C_2$ are the fit constants. Note that each set of windows has a different retardation due to different stresses present in the windows, thus each set has to be calibrated separately. These effects do not change during the course of an experiment, so they are held constant at these values for the duration of the run. These values do vary from run to run due to slight differences in clamping stresses of the chamber, thus they must be calibrated individually for each run. Because these stresses have an amplified effect at shorter wavelengths, we lower bound the wavelength of our fits to 250 nm to eliminate extra errors introduced by these window stresses.

The final hydride optical properties are also fit using a Kramers Kronig consistent B-spline to the measured data of the 4 angles recorded after the dynamic run (including the set delta offsets found above). In some experiments, there can be unusually high stresses in one set of windows causing a distortion in the data. In these cases, the data from that angle is excluded in the model. The surface roughness for the hydride is input into the model, and the Pd thickness found in the pure metal fit is held constant. The authors note that there is lattice expansion from hydrogenation that would cause the $PdH_x$ thickness to be slightly thicker (~4% for unbounded Pd) than the original Pd capping layer, however this is within the error bars for the thickness determination.[52] The experimental $PdH_x$ data for the Pd sample run is used for the capping layer optical properties. Lastly, the thin Mg layer required that the back Cr/Au substrate also be measured and modelled before applying it to the full model for the $Mg/MgH_x$ stack.

The dynamic fit uses two Bruggeman effective medium approximations (EMA): one for the metal under investigation and one for the Pd capping layer. The two materials input into each EMA are the pure metal model and the hydrided model found above. The three fit parameters in the model are percent of the hydrogenated metal for the metal under investigation and the capping layer, and the surface roughness, which is bounded by the defined roughnesses of the metal and the hydride.

*Loading Measurement*
In a separate environmental chamber, the second QCM sample from the deposition is run under identical environmental conditions to the original sample. The loading is calculated as outlined in Murray *et al*, where the stress of the QCM amongst other extraneous effects are corrected for in the loading calculation.[28] To obtain the loading data of the optically measured sample, we first subtract away the frequency change due to the changes of gas partial pressures in the chamber. We then offset and normalize the recorded frequency change. The baseline frequency before $H_2$ is introduced to the chamber is defined to be zero (offset), and the stabilized frequency after the hydrogen loading is complete is defined to be the calculated loading value from the duplicate sample (normalize). During the Pd hysteresis study, a concurrent hysteresis in stress required that each cycle be normalized independently. The loading cannot be directly calculated in the optical measurement chamber because the stress cannot be properly characterized interferometrically due to the geometry of the setup when taking ellipsometric data.

There were two cases where we had to add extra processing to the loading calculations of the metals: Mg and Pd. For Mg, the loading data was normalized in a slightly different process due to the possibility of a thin MgO layer existing at the beginning of the experimental runs. Once hydrogen was introduced to the chamber, the QCM frequency sharply increased, indicating mass loss from the system, before the usual reduction in frequency due to loading. We expect that this effect is due to a MgO layer that is being reduced upon the introduction of the $H_2$. To account for this effect, we normalized the frequency to the top of this peak, defining it as the baseline value. For Pd, the duplicate sample was run in a $D_2$ environment instead of $H_2$. We added the known conversion factor of 0.07 to the final loading value to convert to the proper $H_2$ loading.[9]

## Acknowledgements

The authors are grateful for financial support from Google LLC and thank Matt Trevithick, David Fork, and Ross Koningstein for contributions to this collaboration and M. S. Leite for additional comments and suggestions. The authors also acknowledge fabrication support from the FabLab at Maryland Nanocenter, machining support from Thomas Weimar in the IREAP Machine Shop, and ellipsometric modelling support from J.A. Woollam Co.